\documentclass[12pt,preprint]{aastex}

\shorttitle{Cold dust in globulars}
\shortauthors{Barmby et al.}

\newcommand{\mic}{~$\mu$m\ }
\newcommand{\mice}{~$\mu$m}

\newcommand{\mjsre}{~MJy~sr$^{-1}$}

\begin{document}

\title{A \textit{Spitzer} search for cold dust within globular clusters}

\author{
Pauline Barmby,\altaffilmark{1,2}
Martha L. Boyer,\altaffilmark{3} 
Charles E. Woodward,\altaffilmark{3}
Robert D. Gehrz,\altaffilmark{3}
Jacco Th. van Loon,\altaffilmark{4}
Giovanni G. Fazio,\altaffilmark{2}
Massimo Marengo,\altaffilmark{2} 
Elisha Polomski\altaffilmark{5}
}

\altaffiltext{1}{Department of Physics \& Astronomy, University
of Western Ontario, London, ON N6A 3K7, Canada; e-mail: pbarmby@uwo.ca}
\altaffiltext{2}{Harvard-Smithsonian Center for Astrophysics, 60 Garden Street, Cambridge, MA 02138}
\altaffiltext{3}{Department of Astronomy, School of Physics and Astronomy, University of Minnesota, 
 116 Church St, SE, Minneapolis, MN 55455}
\altaffiltext{4}{Astrophysics Group, Lennard Jones Laboratories, Keele University, Staffordshire, UK}
\altaffiltext{5}{Department of Physics and Astronomy, University of Wisconsin Stevens Point,
 Stevens Point, WI 54481}

\begin{abstract}
Globular cluster stars evolving off the main sequence are known to lose mass, and
it is expected that some of the lost material should remain within the
cluster as an intracluster medium (ICM). Most attempts to detect such an ICM
have been unsuccessful. The Multiband Imaging Photometer for Spitzer 
on the \textit{Spitzer Space Telescope} was used to observe eight Galactic globular 
clusters in an attempt to detect the thermal emission from ICM dust. 
Most clusters do not have significant detections at 70\mice; 
one cluster, NGC~6341, has tentative evidence for the presence of dust,
but 90\mic observations do not confirm the detection.
Individual 70\mic point sources which appear in several of the cluster images
are likely to be background galaxies. 
The inferred dust mass and upper limits
are $< 4\times 10^{-4}M_\sun$, well below expectations
for cluster dust production from mass loss in red and asymptotic giant branch stars.
This implies that either globular cluster dust
production is less efficient, or that ICM removal or dust destruction is more efficient, than
previously believed. 
We explore several possibilities for ICM removal and conclude that present data
do not yet permit us to distinguish between them.
\end{abstract}


\section{Introduction}

Post-main-sequence mass loss affects both present and future
generations of stars. It determines the remnant state of 
stars of intermediate mass and below, as well as affecting the 
state and contents of the interstellar medium out of which new
stars are formed. However, the rate and amount of mass lost by 
red giant branch (RGB) and asymptotic giant branch (AGB) stars
is a major uncertainty in stellar evolutionary theory.
As \citet{origlia07} and \citet{catelan00} point out,
existing formulations rely on empirical laws based on
observations of Population~I stars. There is little theoretical guidance on 
how mass loss should be included in stellar evolutionary models,
or on how it affects population synthesis models based
on the stellar models.

Globular clusters (GCs) have long been the prime testing grounds for stellar evolution
theory. With large populations of low-mass stars of the same age, metallicity,
and distance, they should be ideal places to test ideas about mass loss. 
There are several observational confirmations of the theoretical
prediction that stars more massive than $0.8M_{\sun}$ lose a significant
fraction of their mass after leaving the main sequence. These include
initial-final mass estimates of white dwarfs \citep[e.g.][]{kalirai08} 
and evidence from line profiles of red giants \citep[e.g.,][]{meszaros08}.
To explain the horizontal branch it was argued
\citep[e.g.,][]{rood73} that much of this mass is shed on the RGB, but as AGB
stars, post-AGB stars, and PNe are also observed in GCs \citep{alves00,vanloon07}
some mass must be lost on the AGB as well.
There is also  ample observational evidence that many GCs have 
stars which are currently losing mass, at rates of $10^{-8}-10^{-6} M_{\sun}$~yr$^{-1}$ 
\citep{rj01, ori02, boyer06, origlia07, ita07}. 
From these observations it is difficult to obtain an overall picture of 
time-averaged mass loss rates and duty cycles and their relation to cluster properties
such as metallicity, since only a few stars at a time lose mass. 

A more global
picture of mass loss within GCs might be provided by the intra-cluster medium
(ICM). Mass lost from GC stars is expected to accumulate in the ICM until the cluster
passes through the Galactic disk, when the ICM will be stripped by ram pressure. The ICM
is expected to be mostly gaseous (although the ionization state is unknown),
with a small amount of dust formed in the extended atmospheres of red giants.
The dust should be optically thin and heated by starlight to 50--80~K \citep{ang82,forte02};
its thermal emission should peak in the mid-to-far-infrared  (IR) and 
could be detected as an IR excess at these wavelengths.
Although GC stars are clearly losing mass, detections of the ICM have been
surprisingly difficult. Searches for intra-cluster gas have been conducted for several
decades \citep{bhb83,roberts88,faulkner91,freire01,vl06}; searches for dust 
\citep{lr90,kgc95,ori96,peo97,hop99, matsunaga08} are more recent but have yielded similar results,
mostly upper limits. 
Only one Galactic GC,\object{NGC 7078} (M15), shows clear evidence for an 
infrared (IR) excess \citep{evans03,boyer06}. 
Diffuse infrared emission near \object{NGC 6402} may be associated with the cluster or
with Galactic cirrus, while \object{NGC 5024} has an infrared point source
which may be intracluster dust \citep{matsunaga08}.  An additional cluster, \object{NGC 6356},
has a tentative detection of dust at submillimeter wavelengths \citep{hop98}. 
All of the detections imply dust masses well below the expected levels.

This paper presents results of a \textit{Spitzer}  \citep{sst, gehrz07} survey of eight
Milky Way GCs using 24 and 70\mic observations with the
Multiband Imaging Photometer for \textit{Spitzer} \citep[MIPS;][]{mips}. We describe
our observational strategy and data analysis techniques in \S 2. Section
3 introduces our results, including the detection
of excess IR emission in several of the target clusters and upper limits on
the remaining clusters. Section 4 explores the implication of these new mid-IR results,
including a comparison with predicted dust masses. We conclude in \S 5 
with a summary and suggested follow-up observations.

\section{Observations and data reduction
\label{sec:obs}}

The target list (Table~\ref{tab:targets})
comprises the Milky Way globular clusters that we believed to be the best candidates
for ICM dust detection.
These clusters are reasonably nearby, at high galactic latitude (to reduce
confusing fore- and back-grounds), and have high escape velocities. They are
among the most massive Galactic GCs.
More massive clusters are more likely to harbor a larger number of stars in
the short-lived superwind phase: if the ICM is being removed
from the cluster on short time scales, then we are more likely to
see it in a more massive cluster where the ICM material is being
replenished more quickly.  Clusters with escape velocities
that are much larger than AGB wind speeds ($\sim 10$~km~s$^{-1}$) will also be more
likely to retain ICM material.
At the time this program was planned, all
had existing or scheduled \textit{Spitzer} mid-IR observations
with the InfraRed Array Camera \citep[IRAC;][]{irac}.
Observing clusters with a range of times since last disk passage is desirable, but we did not use this
as a selection criterion because orbital information is
only available for a fraction of Milky Way globular clusters \citep[e.g., from][]{ode97}.
The clusters span a factor of thirty in metallicity (1.5~dex), important for testing
the effects of metallicity on dust production and retention.
All of the clusters are detected with IRAS \citep{kgc95}, and both NGC~104 (47 Tuc) 
and NGC~362 contain stars with known circumstellar dust \citep{ori02,ita07}.

\subsection{IRAC observations
\label{subsec:iracobs}}

IRAC photometry traces the Rayleigh-Jeans tail of the spectral
energy distribution of globular cluster stars. Although cold 
intracluster dust is not expected to radiate at IRAC wavelengths,
understanding the near-IR SED is very useful for quantifying
any IR excess from the dust.
IRAC observations of the clusters were made over the period 2005 July -- 2006 September 
as part of \textit{Spitzer} program identification  (PID) 20298 (PI R. Rood), except for
NGC~5904 which was part of PID 3256 (PI P. Goudfrooij).
Dataset identifiers are given in Table~\ref{tab:iracphot}.
All but one of the clusters were observed in a single IRAC field of view ($5\arcmin\times5\arcmin$) in
all four channels, with 90--252 frames (12~s each) per position
(30 frames, 30~s each for NGC~5904). NGC~104 was
observed in a $5\arcmin\times15\arcmin$ map.
NGC~104 and NGC~6752 were observed in the high-dynamic-range (HDR) mode
which includes short (0.6~s) frames taken before the 12~s frames.
The half-light radius of NGC~104 (190\arcsec) is slightly too large to be
fully covered by the PID 20298 observations. Fortunately, observations
of this cluster covering a wider field of view were available, taken as part of
instrument commissioning in 2003 November (PID 623). 
These observations were taken in 12-second HDR mode with 3 frames
per position, and cover a $25\arcmin\times30\arcmin$ area.

For all clusters, the \textit{Spitzer} Science Center (SSC) pipeline version S14 post-BCD mosaics were used 
for integrated photometry. IRAF\footnote{IRAF is distributed by the National Optical Astronomy Observatories,
 which are operated by the Association of Universities for Research
 in Astronomy, Inc., under cooperative agreement with the National
Science Foundation.}
\textit{apphot} was used to measure integrated magnitudes within 
the half-light radii given in Table~\ref{tab:targets}. For each cluster
the background was measured in an annulus with inner and outer radii
108 and 117\arcsec, respectively, and subtracted from the integrated magnitudes.
For NGC~104 and NGC~6752, the
mosaics made from the short-exposure frames were used
to avoid bright star saturation; 
the other clusters had no saturation problems.
The magnitudes measured in the half-light radii were 
converted to total flux densities with multiplication by 2, and corrected
with application of the `extended source calibration' given by the SSC. 
The results are given in Table~\ref{tab:iracphot}, with electronic links to the observational data.
The uncertainties
in the flux densities due to Poisson noise from the cluster and sky background noise
are very small ($<0.5$\%); uncertainties in the 
absolute \citep[a few percent;][]{reach05} and extended source calibrations
(10\%) dominate the total flux measurements.

\subsection{MIPS observations
\label{subsec:mipsobs}}

The MIPS Photometry Astronomical Observing Template (AOT) was used
to produce $5\arcmin\times5\arcmin$  images of the clusters at 24 and 70\mice.
This field size is comparable to the cluster sizes
\citep[half-light radii from][$<3$\arcmin]{harris96}.
Although a somewhat larger field of view would have been desirable,
this would have substantially increased the observing time.
To set exposure times, we predicted the 24 and 70\mic stellar fluxes
from the clusters' $K_s$-band fluxes \citep[measured from 2MASS images;][]{nan06}
assuming a Rayleigh-Jeans $f_{\nu} \propto {\nu}^{2}$ spectrum.
The 24\mic observations were required to characterize the cluster SED, so
high-significance detections of individual stars were not required. The estimated
24\mic flux was used to compute the expected average 24\mic surface brightness inside the
clusters' half-light radii. The resulting values, 0.8--4.2\mjsre, were expected to be
detectable in 1 cycle of the Photometry AOT with 10 second frame times.
The intra-cluster dust is expected to emit most strongly at 70\mice; however, even at this long
wavelength, the integrated flux of the cluster stars is much brighter than the MIPS detection limits.
We therefore targeted the exposure times to have strong  detections of the
integrated stellar photospheric flux; any dust signal would appear as an IR excess at 70\mice. 
The resulting predicted 70\mic fluxes ranged from 6--150 mJy, requiring 1--10 cycles of the 
MIPS Photometry AOT with 10 second frame times.

\begin{figure}
\plotone{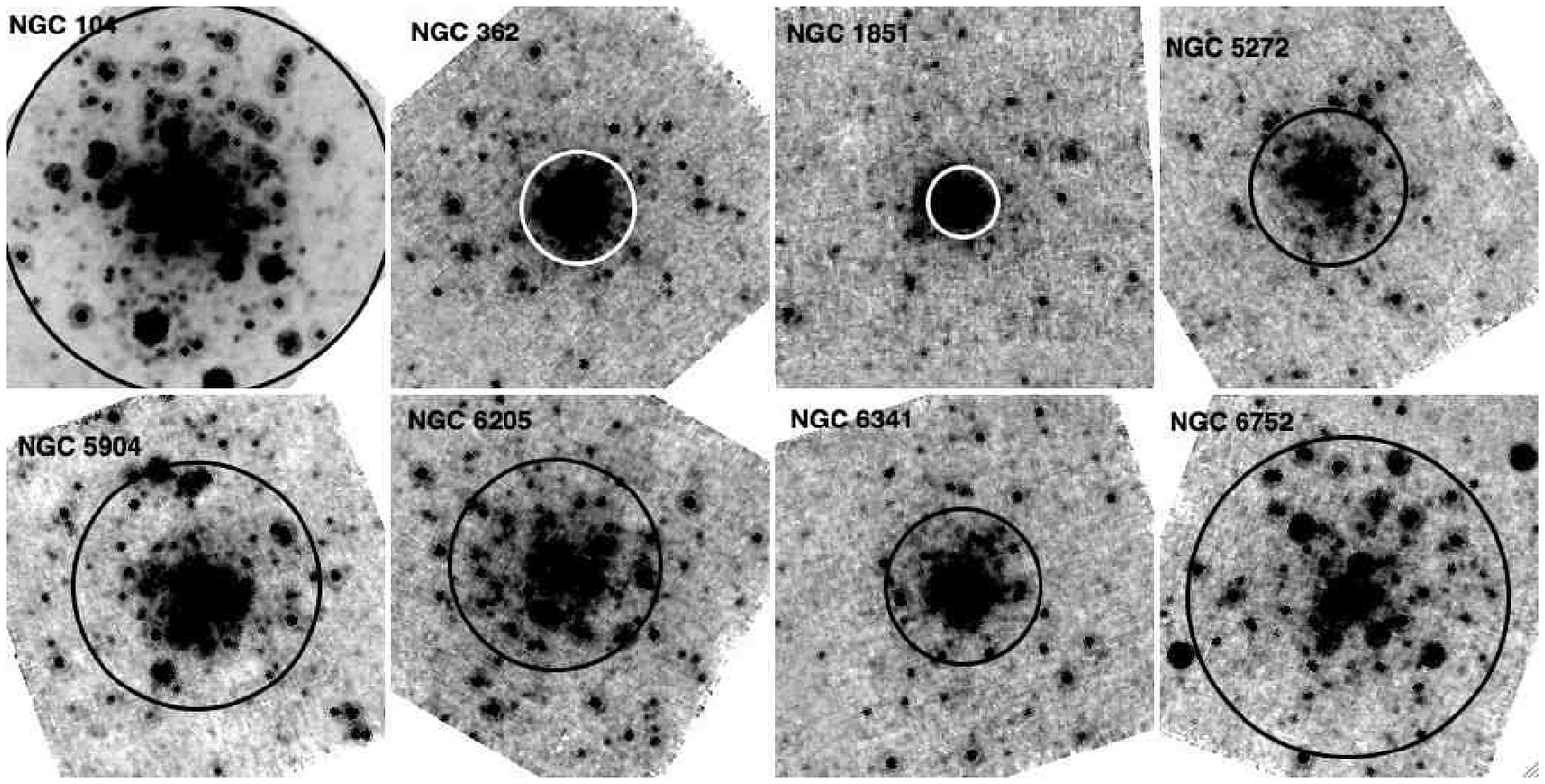}
\caption{MIPS 24\mic images of Galactic globular clusters.
Each image is 5.2\arcmin\ across, with north up and east to the left.
The black or white circles show the half-light radii.
\label{fig:images24}}
\end{figure}

\begin{figure}
\plotone{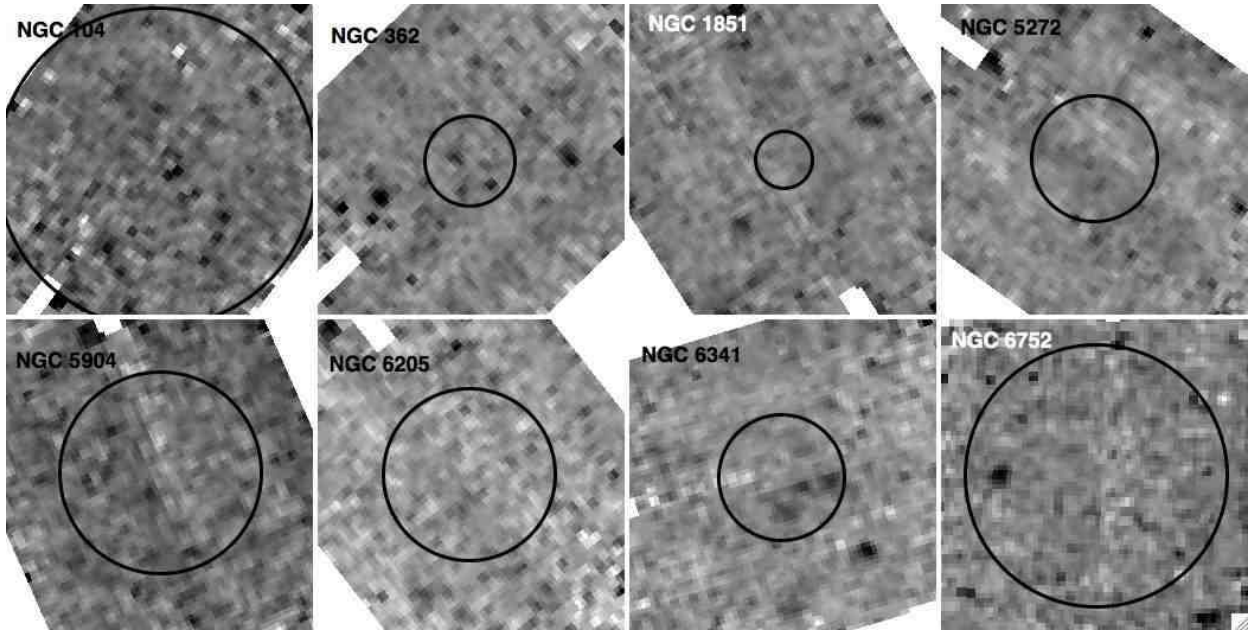}
\caption{MIPS 70\mic images of Galactic globular clusters.
Each image is 5.2\arcmin\ across, with north up and east to the left.
The black circles show the half-light radii (values given in Table~\ref{tab:targets})
\label{fig:images70}}
\end{figure}

The observations were made over the period 2006 June -- 2006 November under \textit{Spitzer} 
PID 30031, a Guaranteed Time program of G.\ Fazio. 
The data were processed with the S14 version of the MIPS
pipeline. Post-pipeline processing for the 24\mic BCD images included `self-calibration' 
(dividing by the normalized median image of `off-source' frames) and mosaicking;
the 70\mic BCD images were time- and column-filtered and then mosaicked. All mosaicking
used the MOPEX program \citep{mk05} with the standard parameters recommended in SSC
documentation. The 24\mic mosaics have a pixel scale of 2.5\arcsec, and the 70\mic mosaics
have a scale of 5.0\arcsec.
Figures~\ref{fig:images24} and \ref{fig:images70} show the cluster mosaic images.
All of the clusters were detected at high signal-to-noise in the 24\mic
images. The same is not true at 70\mice, and the major result of this paper is the
lack of any strong dust signal from any of the eight clusters.

As for IRAC, flux densities in the MIPS bands were measured with aperture photometry.
Inspection of Figure~\ref{fig:images24} shows that the mid-infrared light for several
clusters (NGC~5272 and NGC~6205) appears to have slightly different centers from the
optical centers given by \citet{harris96}. We computed new centers by smoothing
the 24\mic images and computing the flux centroid of the smoothed images in a box of width 4 times
the smoothed FWHM, centered on the optical center. The mid-IR centers are given 
in Table~\ref{tab:mipsphot}. Compared to the IRAC data, the choice of aperture for the
MIPS images is less obvious: the intracluster medium seems more likely to 
sink to the center of the cluster potential well than to follow the
same spatial distribution as the cluster stars. 
[However, we note that both the extended emission detected in NGC~7078 by \citet{boyer06} 
and the possible intracluster dust in NGC~6024 reported by \citet{matsunaga08} are
offset from the cluster centers.]
We decided to use two apertures for each cluster:
the optical half-light radius as for the IRAC data, and a smaller `core' radius,
to maximize signal-to-noise for detection of dust at the cluster center.
The half-light radii are given in Table~\ref{tab:targets}; for these apertures,
background annuli with inner and outer radii 150 and 175\arcsec\ were used.
Following \citet{carpenter08}, we chose a core aperture of 16\arcsec\ radius
(diameter twice the FWHM of the PSF), with background annuli 120--140\arcsec\ in radius.
To allow for subtraction of stellar contribution to the 70\mic signal (see \S\ref{sec:nsflux}),
the 24\mic photometry in this aperture was measured on versions of the images 
convolved to the 70\mic resolution using a kernel described in \citet{gordon08}.

The results of the aperture photometry are listed in Table~\ref{tab:mipsphot}, 
with electronic links to the observational data;
no color  corrections have been made to these values. The only aperture correction made
is a multiplicative factor of 1.13 applied to the large-aperture 70\mic photometry.
Again following \citet{carpenter08}, uncertainties in the MIPS flux density were calculated as 
\begin{equation}
\sigma = \eta_{\rm corr} \eta_{\rm sky}  \sigma_{\rm sky} \sqrt{N_{\rm ap} + N_{\rm ap}^2/N_{\rm sky}}
\end{equation}
where  
$\eta_{\rm corr}$ accounts for the noise correlation between pixels due to re-sampling
during mosaicing ($\eta_{\rm corr} =1$ for 24\mic and 2 for 70\mice),
$\eta_{\rm sky}$ is discussed below,
$\sigma_{\rm sky}$ is the standard deviation of the pixels in the background annulus, and 
$N_{\rm ap}$ and $N_{\rm sky}$ are the number of pixels in the photometry and 
background apertures, respectively. (Poisson noise from the clusters is
negligable compared to the background noise.)
After examining noise in their 70\mic  background and photometry apertures,
\citet{carpenter08} concluded that `excess noise' corresponding to $\eta_{\rm sky}=1.5$ 
was present in their photometry apertures, a conclusion validated by their distribution of
signal-to-noise ratios. Although this precise value may not apply to our
observations, which were made in a slightly different manner, we adopt it here
to be conservative.
The uncertainties in Table~\ref{tab:mipsphot} do not include uncertainties in the absolute calibration 
\citep[2\% at 24\mic and 5\% at 70\mice;][]{gordon07,engelbracht07}. 

Confusion noise due to extragalactic sources is
not expected to affect these observations: \citet{frayer06} derived
the 70\mic confusion limit due to extragalactic sources as 0.3~mJy,
well below the flux regime of the observations presented here.  
Confusion due to Galactic cirrus, however, is an issue:
\citet{jeong05} computed the MIPS 70\mic point source detection limits 
due to Galactic cirrus confusion  at the latitudes of our clusters ($|b| > 27$)
as $0.1-0.2$~mJy, again below our detection limits.
For larger apertures, however, the confusion noise and detection limits scale as 
$d^{1-\alpha/2}$ where $d$ is the aperture size and $\alpha \approx-3.5$ is
the cirrus power spectrum index. The scaling factors range from about
6 to several hundred for the half-light radius apertures used, and hence
cirrus confusion will affect our ability to detect extended emission for
clusters with large angular sizes.

\section{Analysis
\label{sec:results}}

\subsection{Cluster spectral energy distributions}
Figure~\ref{fig:seds} shows the cluster spectral energy distributions (SEDs) from 0.3 to 100\mice.
The new IRAC measurements from Table~\ref{tab:iracphot} and half-light radius MIPS measurements
from Table~\ref{tab:mipsphot} (multiplied by 2 to convert to total fluxes, even though this
may not be appropriate for non-stellar emission) are shown as filled symbols.
Also plotted are extinction-corrected total optical and near-IR magnitudes
given by \citet{harris96} and \citet{cohen07}. IRAS measurements from the Faint Source Catalog 
\citep{kgc95} are also shown; the 12 and 25\mic measurements have been corrected for the cluster extent
following the procedure used by \citet{kgc95} and flux densities in all IRAS bands have been color-corrected using
the values for a 5000~K blackbody given in the \textit{IRAS Explanatory Supplement}.
Upper limits to the 90\mic flux densities given by \citet{matsunaga08} are shown
for the five clusters in common between the two samples.
Overplotted are black-body SEDs derived from fits to the flux densities
(equally-weighted) at $\lambda <5$\mice. All are consistent with the expected average temperatures for
globular cluster stars (4500--5500~K).

A first check of our measurements is to compare them with  previous work.
The \textit{Spitzer} and IRAS results are mutually consistent except for NGC~6341, 
where the 70\mic data are slightly above the IRAS 60\mic upper limits. 
We suggest that the IRAS limit might be slightly too low if there is 
extended emission from the cluster.
For NGC~104, both the 60 and 70\mic observations are below the predictions
based on the black-body fit, possibly  because of the large spatial extent
of the cluster.
Two of our target clusters were observed with the PHOT instrument on ISO by \citet{hop99}:
using those authors' method to convert surface brightness to flux density, their
derived 70\mic flux densities are $130\pm50$~mJy for NGC~104 and $220\pm90$~mJy for NGC~7252. 
Our measurements for these clusters are consistent with these values.
The 70\mic measurements are generally consistent with the 90\mic upper limits
given by \citet{matsunaga08}, except for NGC~6341, where the 70\mic value 
appears to be too large. The detection for this cluster is at $<3\sigma$
significance and may be spurious.

\begin{figure}
\plotone{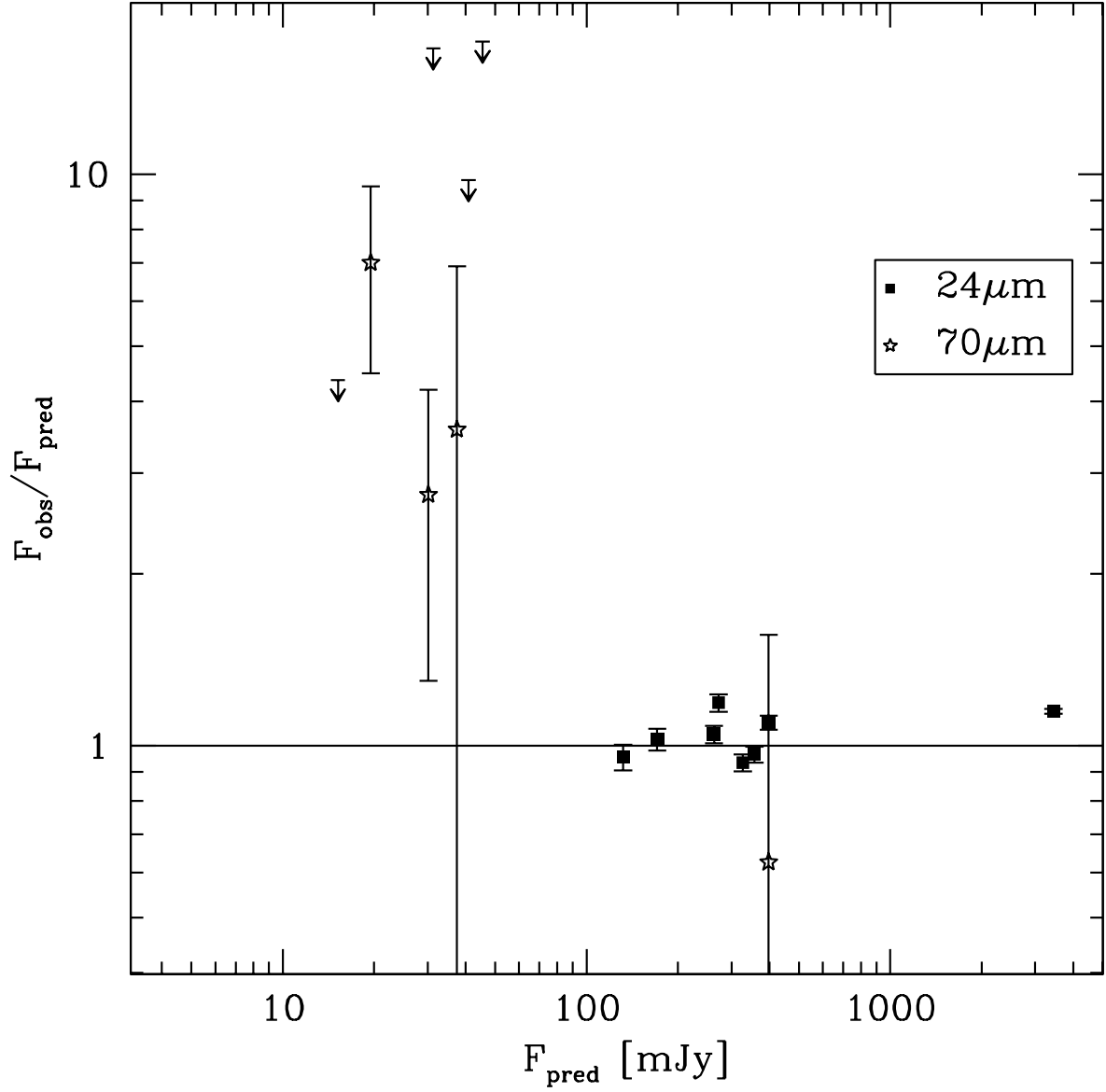}
\caption{Comparison of predicted and measured flux densities for globular clusters
at 24\mic (squares) and 70\mic (stars and upper limits). 
\label{fig:mips_compare}}
\end{figure}

Figure~\ref{fig:mips_compare} compares the predicted (from the black-body fit) 
and measured flux densities (the same values as in the previous figure) at 24 and 70\mice. 
We expected that any ICM dust would likely be too cool to be detectable
at 24\mice; however, circumstellar dust should be warmer and possibly detected
at this wavelength.
All of the 24\mic flux densities are within 20\% of the prediction; six clusters are within
10\%. This is reasonable accuracy considering the difficulties in deriving 
integrated measurements of these resolved objects.
The non-detection of 24\mic excesses in  
integrated photometry may simply reflect the small number of 
mass-losing AGB stars compared to the bulk of the population.
Point-spread-fitting photometry of the 24\mic images, while beyond the scope
of this work, may reveal which stars are surrounded by circumstellar dust.
Seven of the eight clusters in the sample have measurements or upper limits on 70\mic flux
that are more than a factor of 2 above the black-body prediction. 
The exception is NGC~104, for which the 70\mic upper limit is  
{\em lower} than the predicted photospheric value. The problem
could be background subtraction: the half-light radius is
almost as large as the 70\mic image and a ``background'' region is 
difficult to define. 

\subsection{Non-stellar fluxes at 70\mic
\label{sec:nsflux}}

\begin{figure}
\plotone{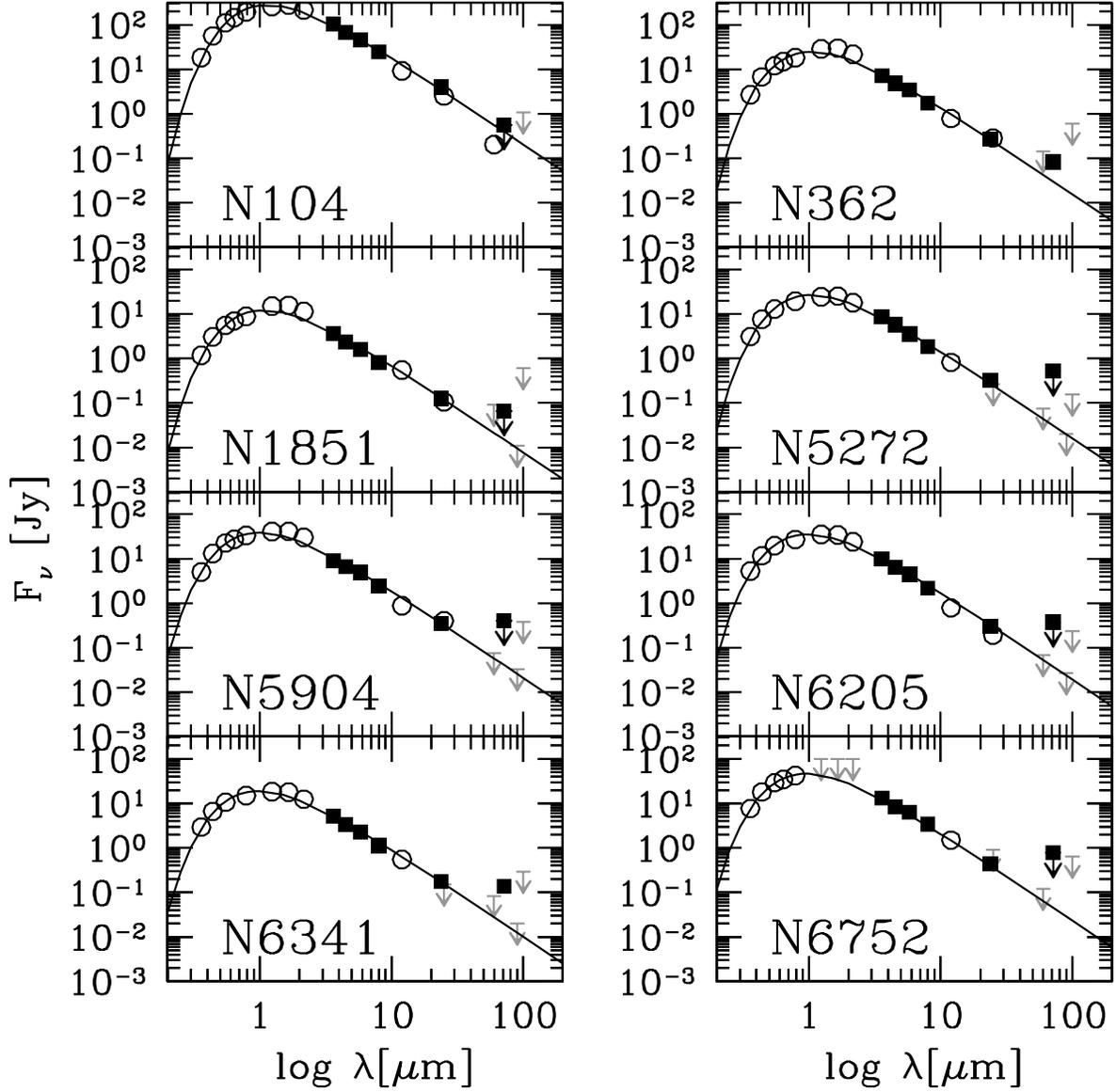}
\caption{Spectral energy distributions of Galactic globular clusters, from
0.3 to 100\mice. Filled symbols are IRAC and MIPS measurements from this work.
Open symbols and grey upper limits are published measurements: optical magnitudes from \citet{harris96},
near-IR magnitudes from \citet{cohen07}, IRAS Faint Source Catalog measurements from \citet{kgc95}, 
and 90\mic upper limits from \citet{matsunaga08}.
\label{fig:seds}}
\end{figure}

Figures~\ref{fig:seds} and \ref{fig:mips_compare} clearly demonstrate that 
only one of the clusters in our sample---NGC~6341---has a significant infrared excess.
To quantify this, we computed (limits on) the 70\mic non-stellar emission
in both the half-light and core apertures. To compute the non-stellar emission
in the half-light aperture, we assumed that any photospheric emission 
at 70\mic follows the optical emission, i.e., half of the total emission
is contained within $R_h$. As an estimate of the stellar emission at 70\mic
we use the predicted value from the black-body fit described above.
Thus the non-stellar emission at 70\mic in the half-light radius is 
\begin{equation}
f_{\rm 70, ns} = f_{\rm 70, obs} - 0.5 \times f_{\rm 70, pred}.
\end{equation}
The fraction of stellar emission contained within the core aperture 
is not as straightforward to compute; to estimate it, we scale the
flux measured at 24\mic in the core aperture (recall that the image
used to measure this flux was convolved to the 70\mic resolution)
by the ratios of the predicted total fluxes at 70 and 24\mice.
Therefore the non-stellar emission at 70\mic in the core aperture is:
\begin{equation}
f_{\rm 70, ns} = 1.76 [f_{\rm 70, obs} - (f_{\rm 24,obs}/f_{\rm 24,pred}) \times f_{\rm 70, pred}]
\end{equation}
where the factor of 1.76 is the aperture correction for a point source measured
in a 16\arcsec\ aperture.
Table~\ref{tab:ns_dust} gives the derived non-stellar fluxes for each
cluster, together with the dust mass predictions and observational limits derived in \S\ref{sec:dust_limit}.

Only one cluster shows even tentative
evidence for the detection of excess IR emission at 70\mice: NGC~6341.  
Figure~\ref{fig:images70} shows that the apparent 70\mic emission in this cluster 
comes from an extended region offset from the cluster core by roughly 43~arcsec, or about $(2/3)R_h$. 
Within the emission region there are three separate peaks which contain about 
60\% of the total flux; it is unclear whether these peaks are separate point sources
(none has a co-located 24\mic source) or merely represent structure in the emission.
In comparison, the dust emission detected in NGC~7078 by \citet{boyer06} is much rounder
and is located closer to the center of that cluster ($\sim R_h/4$); however, the
possible dust cloud in NGC~5024 detected by \citet{matsunaga08} is also far
from the cluster center, at $0.8R_h$.
The IRIS 60\mic sky maps do show some nebulosity near NGC~6341, so it is possible
that the detected flux is in fact not associated with the cluster; however NGC~1851,
at a similar Galactic latitude, also has some nearby nebulosity and does not have a 70\mic detection.
Future data from the Akari All-Sky Survey should have sufficient spatial resolution to 
address the issue of whether there is diffuse emission near NGC~6341.
The lack of any excess infrared emission in the 90\mic Akari observations of NGC~6341 
by \citet{matsunaga08} (see Fig.~\ref{fig:seds}) casts doubt on the reality of our
low-significance detection; however, we retain it for analysis in the remainder of this paper.

\subsection{Point sources detected at 70\mic
\label{sec:ptsrc70}}

While only some of the clusters have spatially diffuse 
IR emission within $R_h$, all of the 70\mic images except those 
of NGC~104 and NGC~5904 do contain some faint point sources
which are not concentrated toward the cluster centers. 
[Most of the point sources identified by \citet{matsunaga08} are off the fields of view of our images.]
What is the nature of these sources? The 70\mic data provide only marginal detections
and poor spatial localization, but the 24\mic data provide additional
clues to the nature of the sources. 

Initially, visual examination was used to  generate
lists of possible point sources in each cluster image. Using an estimated 70\mic position 
as a starting point, we performed aperture photometry on the 24 and 70\mic images
using IRAF/phot, with the photometry routine allowed to re-center the sources. 
Several 70\mic sources were off the edges of the 24\mic
images, or not visible at the shorter wavelength: the latter group could
either be very red sources, or noise fluctuations in the 70\mic images.
Standard photometric apertures as
given in the MIPS Data Handbook were used (at 70\mice, aperture
radius 18\arcsec\ and background annulus $39-65$\arcsec; at 24\mice,
aperture radius 6\arcsec\ and background annulus $20-32$\arcsec)
and appropriate aperture corrections were applied. 
We did not attempt IRAC photometry for the 70\mic sources: the IRAC
images of the GCs are significantly more crowded, and we were doubtful about
the reliability of measurements in these regions.
 
\begin{figure}
\plotone{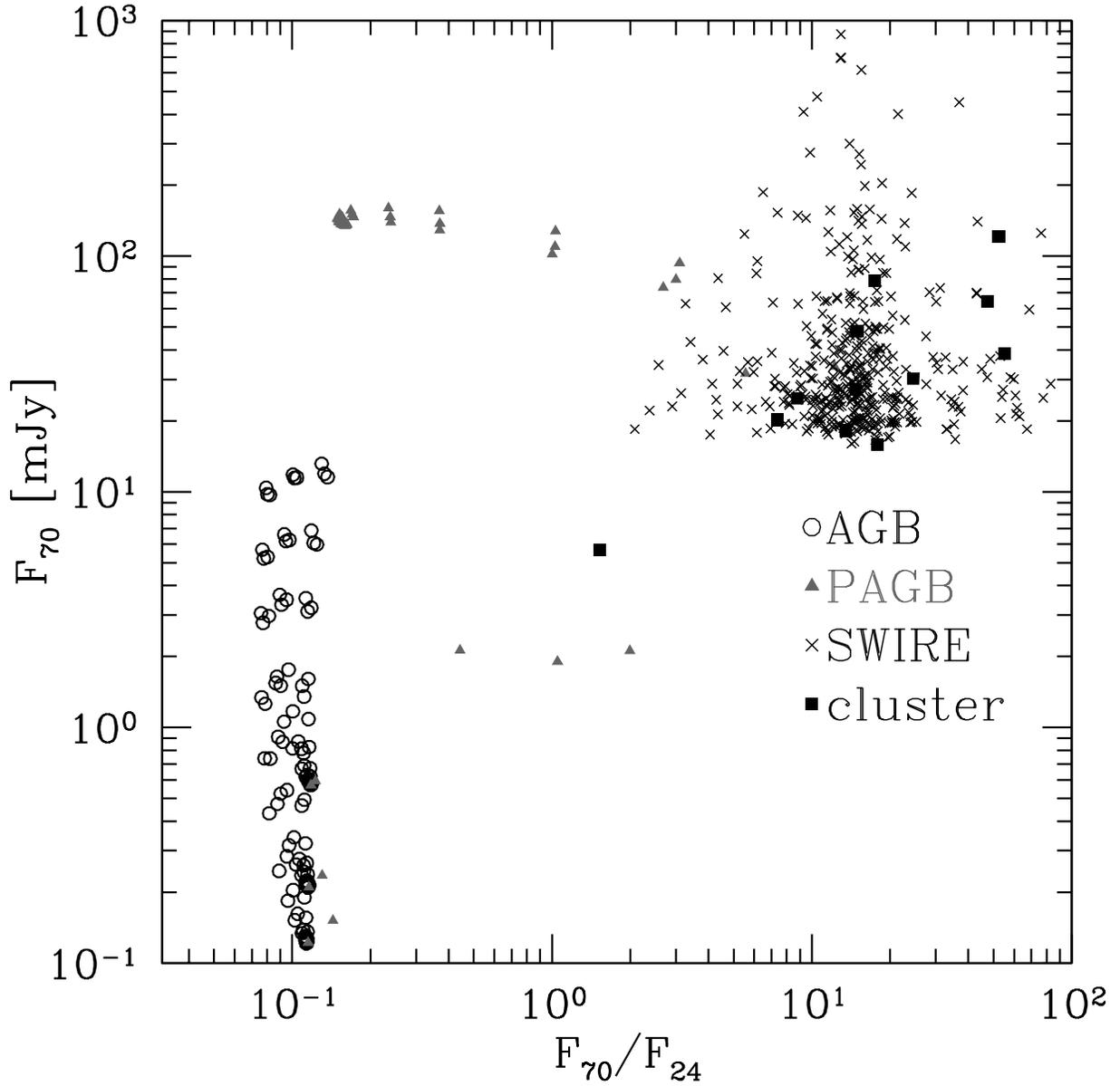}
\caption{MIPS color-magnitude plot for sources detected at 70\mic (squares),
AGB star (open circles) and post-AGB star (filled triangles) models from \citet{gro06},
and SWIRE ELAIS-N2 sources (crosses).
The AGB models and cluster source fluxes have been scaled to a distance of 8.5~kpc, assuming the
sources to be at the distance of the cluster.
\label{fig:mips_cmd}}
\end{figure}

The 70\mic sources in the globular cluster fields could be 
stars within the cluster, unrelated Galactic stars, or background
galaxies. To discriminate between these possibilities, a color-magnitude diagram 
was made from the MIPS photometry, shown in Figure~\ref{fig:mips_cmd}. 
For comparison, the AGB and post-AGB models 
(restricted to oxygen-rich models with $\dot{M}\le 10^{-6}M_{\sun}$, which are
most appropriate for comparison with GCs) of \citet{gro06} are also shown.
The model fluxes have been scaled down by a factor of 5 to
reflect a dust-to-gas ratio appropriate for GC metallicities.
The model fluxes are scaled to a common fiducial distance (8.5~kpc) 
so the same practice was followed for the cluster sources. In addition to the 
evolved star models, we show the (unscaled) flux densities and ratios for
70\mic  sources in the ELAIS-N2 field of the SWIRE Legacy
Survey \citep{swire}; these sources are predominantly extragalactic.
 All but one of the cluster field 70\mic sources are very red, with typical values of
$f_{70}/f_{24}\approx 10$, compared to the Rayleigh-Jeans value of $\sim0.1$.
The cluster source colors are more similar to those of the galaxies than to most
of the evolved-star models, and the stellar models with similar colors are significantly brighter.

From the colors, we tentatively conclude that the 70\mic sources in the cluster fields
are likely background galaxies. 
The high Galactic latitude of the clusters ($|b|>26\arcdeg$) supports this 
conclusion, as do the results of \citet{boyer08} from their study
of 70\mic sources near $\omega$~Cen and those of \citet{matsunaga08}.
The number density of sources is also consistent
with this conclusion: there are 23 sources in a total of 346 arcmin$^2$ for a density 
of 0.066 arcmin$^{-2}$. The number counts given by \citet{frayer06} for the GOODS-North field 
show that there are 0.07 galaxies~arcmin$^{-2}$ to flux limits of 11~mJy 
(appropriate for our average field), comparable to the number in the cluster fields.
\citet{ita07} observed NGC~104 and 362 at mid-infrared wavelengths from 2.4--24\mice.
Those authors found eight red sources ($F_{24}/F_7>1$) in NGC~362; no such sources were
found in NGC~104. Comparing our images of NGC~362 to those of \citet{ita07}, we find that their
red sources A, D and F are within our field of view. Only source D  is detected at 70\mic;
it has $F_{70}/F_{24}\approx15$, consistent with the colors of a galaxy.

\section{Discussion}
\subsection{Dust masses and limits
\label{sec:dust_limit}}

The non-stellar 70\mic fluxes computed in \S\ref{sec:nsflux}
allow us to estimate the amount of dust present in these clusters. 
For the clusters with negative non-stellar fluxes, we computed the
dust mass upper limits  as $3\sigma_{\rm flux}$.
We follow \citet{evans03} in assuming any dust to be optically thin, so the
dust mass is given by:
\begin{equation}
\frac{M_{\rm d}}{M_{\sun}} = 4.79\times 10^{-17} f_{\nu} {\rm[mJy]} \frac{D^2_{\rm kpc}}{{\kappa}_{\nu} B_\nu(T_d)}
\end{equation}
where $\kappa_\nu$ is the dust absorption coefficient
and $B_\nu(T_d)$ the Planck function, both in cgs units.
Our data do not constrain the dust temperature, so we use the value
found by \citet{boyer06} for NGC~7078 ($T_d=70$~K), and the same $\kappa_\nu=56$~cm$^2$~g$^{-1}$.
Different assumptions about these two quantities can change the derived dust
masses by several orders of magnitude; for example, \citet{evans03} estimated the
dust mass in NGC~7078 by using $\kappa$ for amorphous fayalite at 50~K; at 70\mic
this substance has $\kappa= 280$~cm$^2$~g$^{-1}$, or five times the value used by \citet{boyer06}.
On the opposite extreme, \citet{kgc95} assume silicate grains with $T=45$~K;
with $\kappa_{\nu} \propto \nu$,  the absorption coefficient at 70\mic is 36~cm$^2$~g$^{-1}$.
 Table~5 of \citet{hop99} shows the
effects of different assumptions about the dust temperature on derived dust masses:
changing from $T_d=70$~K to $T_d=40$~K, for the same composition, increases
the inferred $M_d$ by about a factor of ten, while increasing from $T_d=70$~K to $T_d=100$~K decreases
$M_d$ by about a factor of 2.5.
Uncertainties in the dust composition and temperature therefore result in uncertainties of $\lesssim 2$
orders of magnitude in derived dust masses.

Table~\ref{tab:ns_dust} gives the dust mass upper limits 
derived from the 70\mic measurements in both the half-light and core apertures.
All values reported are $3\sigma$ upper limits, except for the 
possible ($2.2\sigma$) detection of total dust mass in NGC~6341. This is our only observation resulting in
even a tentative detection; we show it as a value rather than an upper
limit in the table and following plots in order to give a visual estimate of the uncertainties.
Because the cluster emission is generally very faint, we have not attempted
to subtract any point source emission from the total fluxes; if point sources
within the apertures are unrelated to the clusters then the true upper
limits would be even lower.
For all of the clusters in our sample, the limits on the total dust masses are 
$< 4 \times 10^{-4}M_\odot$; the limits on core masses are an order of
magnitude lower, $< 4 \times 10^{-5}M_\odot$. 
\citet{hop99} gave upper limits to the dust masses for NGC~104 and NGC~5272;
our total value for NGC~104 is slightly above their limits while our core value
is slightly below. Our total value for NGC~5272 is comparable to that of
\citet{hop99} for 70~K dust. 
Our value for NGC~5272 is also below that given by \citet{peo97} for
80~K dust based on their observations of this cluster at millimeter wavelengths.
Our upper limits are above those of \citet{matsunaga08} for the five
clusters in common, due to the greater sensitivity of their observations.

The dust mass limits we infer are consistent with measurements
and upper limits on gas masses. Converting dust masses to gas masses
by multiplying by $10^{2-{\rm [Fe/H]}}$, we derive gas masses or upper limits
for our sample clusters in the ranges $0.002-0.07 M_\sun$ (core)
and $0.04-1.4 M_\sun$ (total). 
The only confirmed H~I detection, $0.3 M_\sun$ in NGC~7078 \citep{vl06},
is nicely within this range.
Our inferred limits on the gas mass are
consistent with the results of previous searches for intra-cluster neutral
and ionized hydrogen, as summarized in \citet{roberts88}.
Two clusters have reported detections of  ionized
gas:  NGC~104 \citep{freire01} and NGC~1851 \citep{gl77}.
\citet{freire01} derived a total gas mass of $0.1M_\sun$ and
stated that they expected essentially all the gas in the cluster to be ionized;
\citet{gl77} estimated the total mass of ionized hydrogen in NGC~1851 as $0.02M_\sun$.
Our gas mass upper limit for NGC~104 is about the same as the
\citet{freire01} detection while our upper limit for NGC~1851 is above the
\citet{gl77} level.

\subsection{Expected dust masses}
To understand the significance of the dust masses and upper limits derived 
from the \textit{Spitzer} observations, we need  
theoretical estimates of the expected dust content of the clusters in the sample.
Various estimates of the dust mass within a globular cluster are
possible, but little detailed modeling has been done with the exception of
the work of \citet{ang82}. Those authors combined dynamical models of star clusters
with radiative transfer models to predict dust grain temperatures and
luminosities for a few individual clusters, given the presence of
$10^{-2}M_\odot$ of dust. While it would be valuable to repeat such 
modeling with modern radiative transfer codes and understanding of 
dust properties, for present purposes a simpler estimate of globular cluster
dust content will suffice. Following \citet{tw75}, the dust mass 
may be estimated by:
\begin{equation}
\frac{M_{\rm d}}{M_{\sun}} = \frac{\tau_c}{\tau_{\rm HB}} N_{\rm HB} (\delta M) \frac{10^{\rm [Fe/H]}}{100}
\label{eq:dustpred}
\end{equation}
where $\tau_{\rm HB}=1.8\times10^8$~yr is the horizontal branch lifetime, 
$\tau_c$ is the time since last crossing of the Galactic plane,
$N_{\rm HB}$ is the number of stars on the horizontal branch, 
$(\delta M)=0.2M_\sun$ is the gas mass lost from each star, and 
${\rm [Fe/H]}$ is the cluster metallicity.
The last factor scales for the change in gas-to-dust ratio with metallicity.
Values for $\tau_c$ are listed in Table~\ref{tab:targets}; for the two clusters not listed
by \citet{ode97} we assume a lower limit of $10^6$~yr.
To estimate $N_{\rm HB}$ we use the  `specific evolutionary flux' method of \citet{rb86}
\citep[see also][]{evans03}:
\begin{equation}
N_{\rm HB} = BL_{\rm bol}\tau_{\rm HB}
\end{equation}
assuming $B=2\times10^{-11}$~stars~yr$^{-1}$~L$_\sun^{-1}$ as appropriate for globular clusters, and 
$L_{\rm bol}=2\times L_V$ \citep{rb86} using $M_V$ and $E(B-V)$ given by
\citet{harris96}.
For one cluster, we can compare this dust mass estimate to the current dust production rate:
\citet{origlia07} have recently used {\it Spitzer} data to identify the mass-losing stars
in  NGC~104. Summing their mass loss rates for the brightest stars ($M_{\rm bol}<-2$) 
and converting to a dust production rate with a gas-to-dust ratio of 200, we estimate
${\dot{M}}_d = 1.4\times10^{-8}M_\sun$~yr$^{-1}$. When multiplied by $\tau_c=6\times10^7$~yr,
this yields a total dust content of $0.84 M_\sun$,  about twice that predicted from Eq.~\ref{eq:dustpred}.
The observations of \citet{origlia07} are likely to be affected by blending; correcting for
this would lower the inferred dust mass to a value closer to the prediction.
Even without such a correction, the comparison  gives us confidence that the dust mass predictions  
are of the correct order of magnitude.

\begin{figure}
\plotone{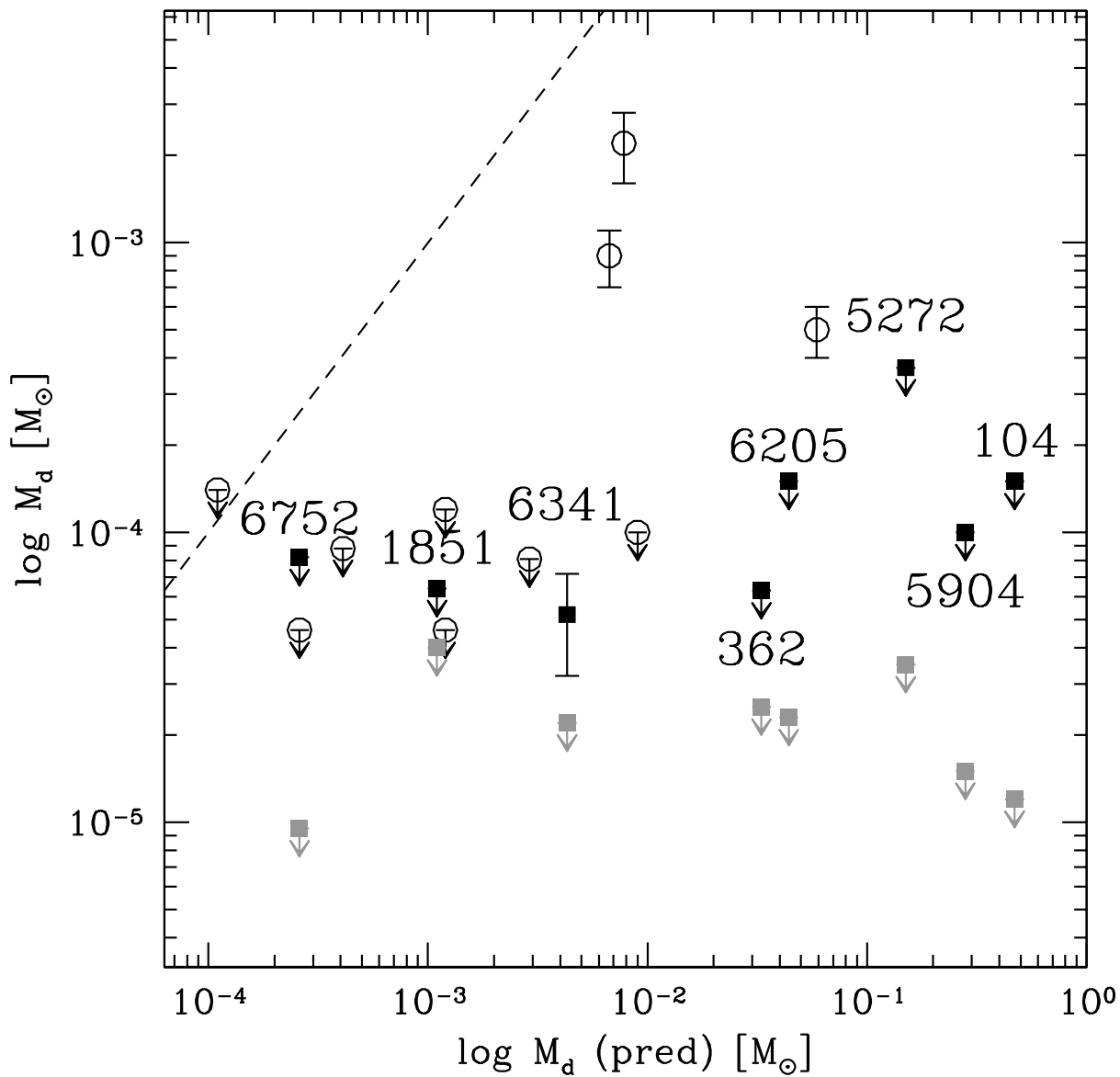}
\caption{Cluster dust masses predicted with Eq.~\ref{eq:dustpred} versus 
derived masses and upper limits from the MIPS observations for both
total dust mass (filled black symbols) and core dust mass (filled grey symbols).
Previously published dust limits are open symbols, not labeled with cluster NGC number. 
Dashed line is the line of equality. 
\label{fig:dustpred}}
\end{figure}

Figure~\ref{fig:dustpred} compares the predicted and observed dust masses and upper limits.
To augment our sample we analyse additional clusters observed by other groups: 
\object{NGC~6356}%
\footnote{For this object we plot the more conservative `dust in beam' mass given 
by \citet{hop98}; \citet{matsunaga08} use the larger `dust in core' value. 
Regardless of this choice, this cluster still has the largest inferred mass.}
\citep{hop98}, \object{NGC~6656} \citep{peo97}, 
and NGC~7078 \citep{boyer06}, and the six other clusters observed by \citet{matsunaga08}: 
\object{NGC 1261}, \object{NGC 1904}, \object{NGC 2808}, \object{NGC 5024}, \object{NGC~5139} 
\object{NGC 5634} and \object{NGC 6402}. 
Where several values for dust mass or upper limits are given in published
works, only the one with assumed dust temperature closest to 70~K is used.
All of our measurements are well below the predicted values; for most clusters the
disagreement is more than an order of magnitude.
The same is true for clusters observed by other groups, in that
only two clusters have the inferred dust mass close to the predicted
value: NGC~6356 \citep{hop98} (only for specific assumptions
about the dust distribution), and NGC~5634 \citep{matsunaga08}, one of the
lowest-mass, lowest-metallicity clusters shown.
While assuming a lower dust temperature or higher absorption coefficient would
increase the dust masses inferred from the observations, it seems unlikely 
that this can explain the three order of magnitude discrepancy for clusters such as NGC~5272.
The other explanation is that one or more of the assumptions involved in
Eq.~\ref{eq:dustpred} is incorrect: what if globular cluster stars lose much less
than $0.2M_\sun$ on average, the number of horizontal branch 
stars does not scale with $L_{\rm bol}$, the dust-to-gas ratio does not scale
with metallicity, or the dust is removed from clusters on a timescale much shorter 
than the disk passage time $\tau_c$?
Direct observations of red giant mass loss and color-magnitude diagram modeling
are consistent with  mass losses of approximately $0.2M_\sun$ \citep{origlia07,caloi08},
and the scaling of $N_{\rm HB}$ with $L_{\rm bol}$ is well-grounded in stellar evolutionary theory.
We address the remaining factors below.

\begin{figure}
\plotone{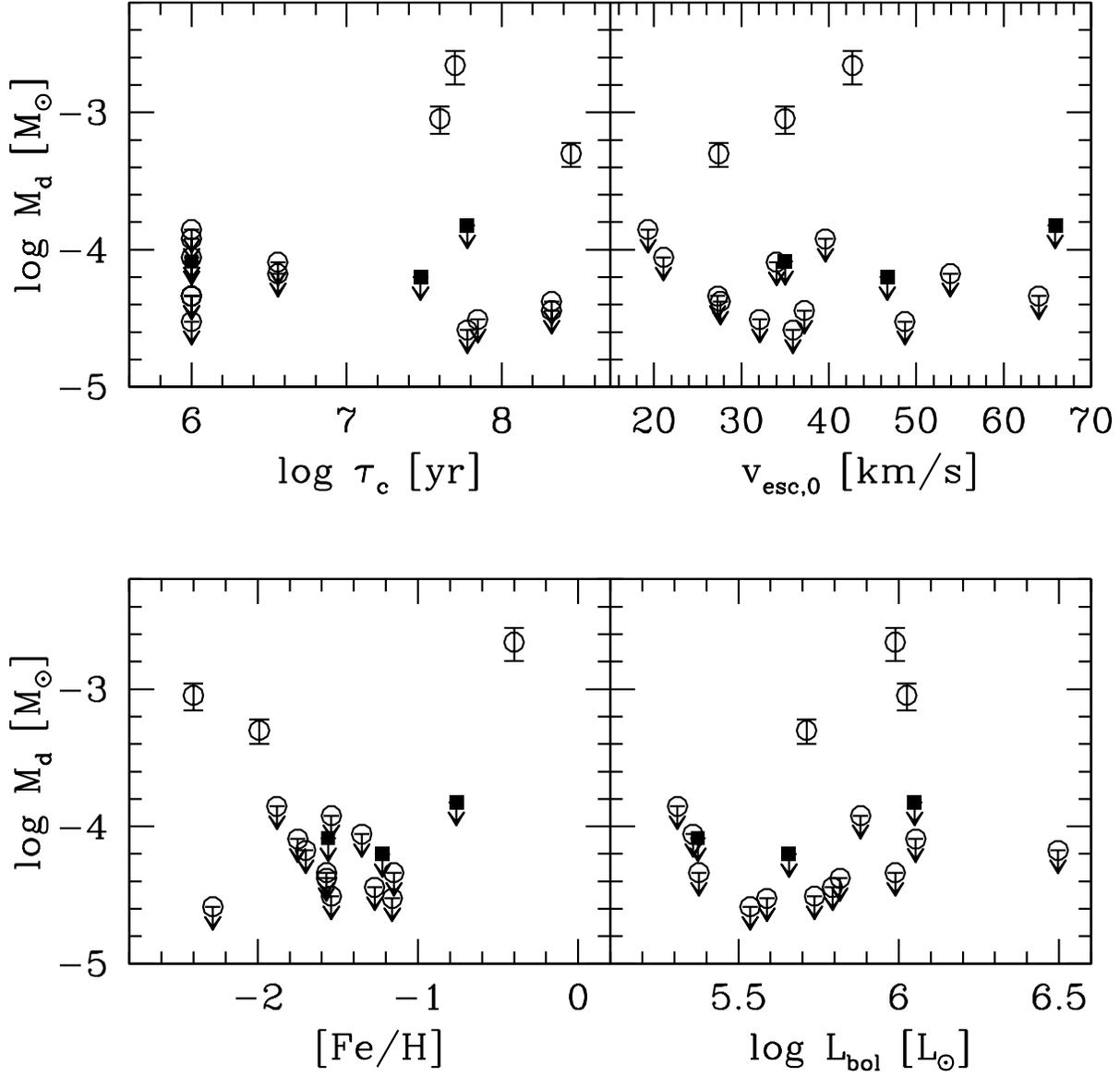}
\caption{Derived cluster dust masses and upper limits versus
time since last Galactic disk passage (upper left), 
central escape velocity (upper right),
metallicity (lower left), and bolometric luminosity (lower right).
Filled symbols: results from this work (total masses), open symbols: previously published work.
An uncertainty of 20\% has been assigned to $M_d$ for NGC~5024 as no value was given in the
original work \citep{matsunaga08}.
\label{fig:masslim}}
\end{figure}

With a small sample of clusters, and an even smaller number of low-significance detections, 
deriving conclusions about the effects of GC properties
on mass loss is difficult.
Figure~\ref{fig:masslim} shows the dust masses and limits plotted as a
function of ${\rm [Fe/H]}$, $L_{\rm bol}$, $\tau_c$, and $v_{\rm esc,0}$.
(In this and the following figure, we plot the more stringent dust mass limits 
of \citet{matsunaga08} for the five clusters in common with our sample.)
There are no clear correlations: while the clusters with large inferred
dust masses tend to have large $\tau_c$, consistent with the expectation
that intracluster dust builds up between cluster passage through the disk, 
other clusters with similar values of $\tau_c$ have much smaller inferred masses. 
This agrees with the finding of
\citet{boyer06} that the ICM dust in NGC~7078 must have been collecting 
for a time $t\ll\tau_c$, and with the computation by \citet{matsunaga08}
that cluster dust lifetimes $\tau_d \ll \tau_c$. 
The clusters with dust detections do not have
the highest central escape velocities, suggesting that possession of
a deep potential well is not in itself enough to guarantee the
presence of an ICM. The clusters cover a relatively 
small range in bolometric luminosity; this
parameter  does not seem to have a strong effect on inferred dust mass.

How does metallicity relate to intracluster dust?
The two clusters with the largest inferred dust masses 
(NGC~7078 and NGC~6356) have the lowest and highest metallicities in the sample.
NGC~5024, with a tentative dust detection, also has a low metallicity.
The globular cluster data do not help to clarify the controversy over the
connection between metallicity and dust production: while some studies
show that metallicity and mass loss are not correlated in carbon-rich AGB stars 
in the Magellanic Clouds \citep{lagadec08, sloan08}, others find that
metal-poor carbon stars, like oxygen-rich giants, form less dust \citep{vanloon08}.
The clusters considered here are almost all more metal-poor
than the Magellanic Clouds, so their AGB stars are expected to be oxygen,
rather than carbon-rich; however the effects (if any) of metallicity on dust mass
loss are not evident from the present data.

\subsection{Dust removal mechanisms}

\begin{figure}
\plotone{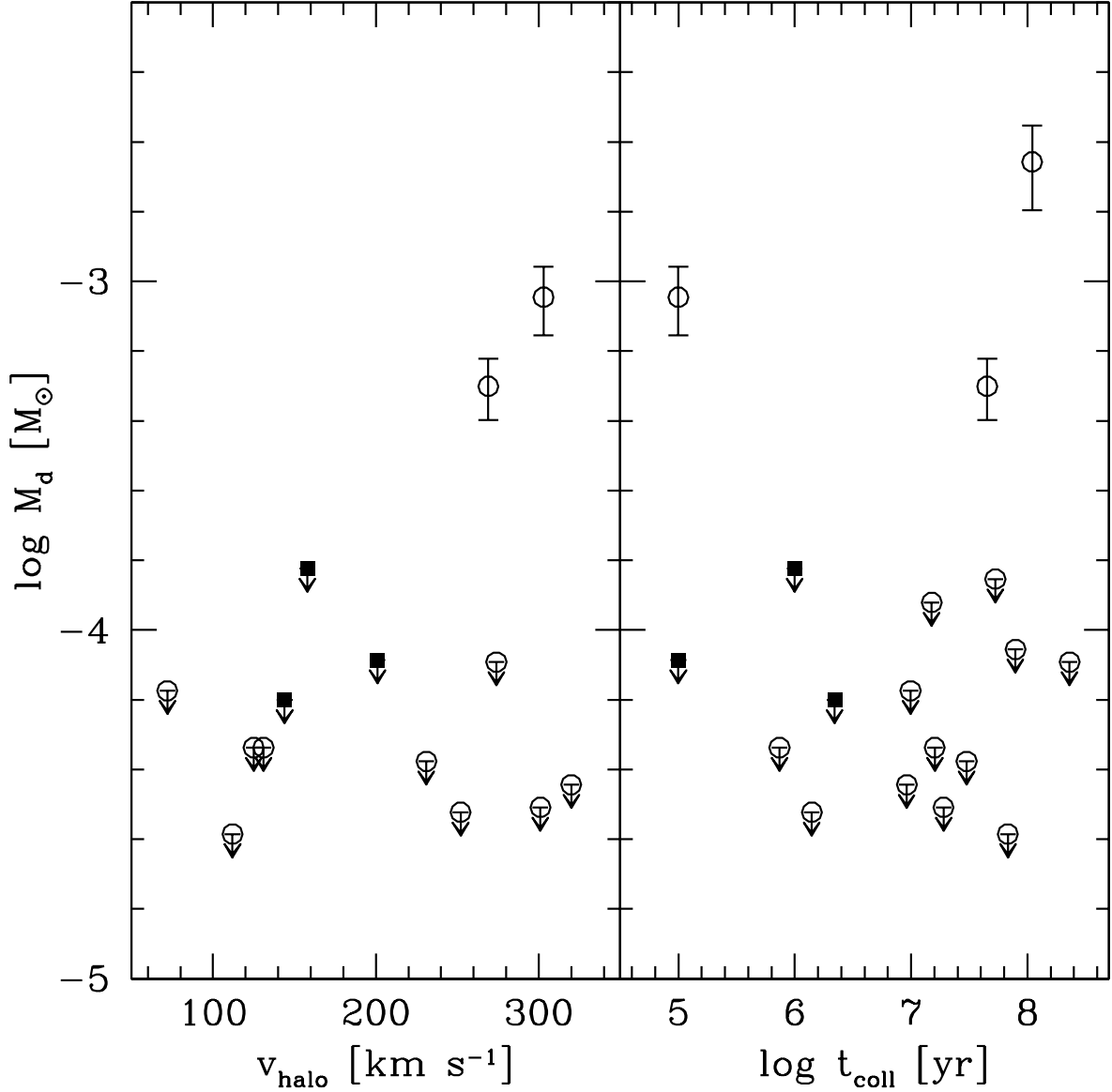}
\caption{Derived cluster dust masses and upper limits versus
velocity relative to the halo (left) and estimated time
between stellar collisions in the core (right).
Symbols as in Fig.~\ref{fig:masslim}.
In the left panel, clusters NGC~1261, 5646, 6356, and 6402
are not shown, as orbital parameters were not available.
In the right panel, both NGC~6752 and NGC~7078 are plotted at
$t_{\rm coll}/N =10^5$~yr; these clusters are core-collapsed
\citep{mv05} and should have short collision times.
\label{fig:dust_remove}}
\end{figure}

Observations of mass-losing stars within GCs show that mass loss is
occurring at roughly the expected rate, so the non-detection of the ICM
means that it must somehow be removed. One possible mechanism is
ram-pressure stripping of the cluster ICM by the Galactic halo. Support for
this mechanism is provided by detections of 
diffuse X--ray emission near several Galactic globular clusters
\citep{okada07,kg95}; \citet{okada07} show that the temperature and luminosity
of the diffuse emission is consistent with it arising from stripped intracluster
gas, and that the energetics require nearly all of the cluster gas to be removed.
Those authors also state, to show diffuse X--ray emission, a cluster should
have a velocity relative to the Galactic halo $v_h>150$~km~s$^{-1}$.
We computed $v_h$ for our sample clusters  from the space velocities
given in \citet{ode97}, \citet{dinescu99}, or \citet{casettidinescu07}
assuming a non-rotating halo;
these are plotted against dust masses and limits in the left panel of Figure~\ref{fig:dust_remove}.
The high velocities of both NGC~7078 and NGC~5024 appear to be problematic for halo stripping;
a measurement of the space velocity of NGC~6356 might help to clarify
the situation.
There is certainly no evidence that clusters with low $v_h$ have more dust;
however, the correlation of cluster kinematics with metallicity combined with any 
dependence on metallicity of the dust mass limits may also confuse the issue.

Another possible ICM removal mechanism is kinetic energy injected by stellar collisions, 
as proposed by \citet{ucr08} for M15. Those authors use an N-body model to estimate the time between
collisions in the cluster core; their result is of the same order of magnitude as a less-sophisticated
estimate based on Eq.~8-125 of \citet{bt87}:
\begin{equation}
\frac{t_{\rm coll}}{N} = [7.9\times10^{14}\thinspace {\rm yr}] \frac{\sigma_c}{\rho_0^2 R_c^3}
\end{equation}
where $N$ is the number of stars within the cluster core (of radius $R_c$~pc), $\sigma_c$ is the
central velocity dispersion in km~s$^{-1}$, and $\rho_0$ is the central mass density
in ${\rm M}_\sun$~pc$^{-3}$. Using the values for $R_c$,  $\rho_0$, 
and $\sigma_c = (2\pi)^{-1/2} rho_0^{1/3}f_0^{-1/3} $, 
from from King-model fits given in \citet{mv05}, we computed $t_{\rm coll}/N$ for each
cluster in the sample; it is plotted against $M_d$ in the right panel of Fig.~\ref{fig:dust_remove}.
Again, there is no clear correlation of the time between collisions and
dust masses: while both NGC~6356 and NGC~5024 have long  $t_{\rm coll}$,
so do many other clusters. NGC~7078 is core-collapsed and should have a short
central collision time.
We conclude that  neither ram pressure stripping and stellar collisions as the cause of ICM removal
can be ruled out by the current data.

\section{Conclusions}
The intracluster medium of globular clusters remains elusive. 
Our sample comprised eight Galactic globular clusters
thought to have the best chance for retaining an intracluster medium.
No significant extended emission is detected from seven of the eight
target clusters: any such emission may well be hidden underneath confusion noise from Galactic cirrus.
One cluster, NGC~6341, has tentative evidence for excess IR emission from
cool dust, although the lack of a 90\mic excess casts doubt on the detection.
All of our inferred dust masses and upper limits are $< 4\times 10^{-4} M_{\odot}$,
compared to predicted dust masses ranging from $10^{-0.8}-10^{-3.6}M_{\odot}$ for the same
clusters. Searches for intracluster gas have been similarly unsuccessful \citep[e.g.,][]{vl06}. 
The lack of intracluster dust and gas
implies that either (1) evolved stars in globular clusters make less dust than predicted, or 
(2) the ICM escapes from the clusters more quickly than expected. 
The first possibility seems unlikely because of the evidence
to the contrary in both NGC~7078 and NGC~5139, where two of the three
dustiest, most mass-losing stars have ${\rm [Fe/H]} < -2.0$ \citep{boyer06,boyer08}.

It is important to point out that ICM escape is not just a trivial occurrence
in the life of a globular cluster: substantial mass loss over a GC's history 
could affect its dynamical evolution  and perhaps even its survival as a bound system \citep{vm06}.
Several mechanisms for ICM escape  have been discussed above,
with energy injected by stellar collisions perhaps being favored over
ram pressure from Galactic halo gas.  Stellar winds provide another
possible mechanism; giant stars with wind velocities above the escape velocity of a typical cluster
have been observed \citep{sds04,dupree92}.
While there is observational evidence that metallicity does
not strongly affect the mass loss rate of individual stars \citep{mv07, sloan08}, 
stellar wind velocities do appear to scale with metallicity \citep{marshall04}.
Faster winds in AGB stars  in metal-rich clusters could therefore
increase the rate at which the ICM is driven out of the cluster.
The existing small sample of clusters does not
permit us to distinguish between the possibilities for ICM removal.

Decades of searching have yielded few detections of the elusive 
globular cluster ICM. A possible strategy for future observations hoping
to detect the ICM is to study the removal mechanisms in detail and
identify those clusters most likely to retain some ICM. 
Observations of distant clusters which have had
few interactions with the Milky Way could place limits on the importance of
halo stripping, although the larger distance of such clusters reduces the
effective sensitivity. While infrared facilities available in the near future 
will not greatly improve on the sensitivity of \textit{Spitzer}, observing a larger sample
of clusters would allow a more thorough exploration of the dust content
as a function of cluster properties.
Observations with future radio facilities such as the Square Kilometer Array
will place more stringent limits on the gas content.
Together with more sophisticated modeling efforts and detailed examination 
of mass loss in individual cluster stars, a multi-wavelength observational
approach provides the best chance for finding and explaining the elusive intracluster
medium in globular clusters.

Facilities: \facility{Spitzer (IRAC, MIPS)}

\acknowledgments
We thank the anonymous referee for a careful reading of the manuscript and helpful comments
and R.T. Rood for early access to the IRAC data on NGC~6205 and NGC~6341.
This work is based on observations made with the \textit{Spitzer Space Telescope},
which is operated by the Jet Propulsion Laboratory, California Institute of 
Technology under NASA contract 1407. 
P.~B. was partially supported by a Discovery Grant from the Natural Sciences
and Engineering Research Council of Canada.
M.~L.~B., C.~E.~W., E.~P., and R.~D.~G. are supported in part by
NASA through \textit{Spitzer} contracts 1276760, 1256406, and 1215746
issued by JPL/Caltech to the University of Minnesota.

\clearpage

\clearpage

\begin{deluxetable}{lrlrrrr}
\tabletypesize{\scriptsize}
\tablecaption{Globular cluster targets\label{tab:targets}}
\tablewidth{0pt}
\tablehead{
\colhead{Target}&\colhead{Distance\tablenotemark{a}}&\colhead{[Fe/H]\tablenotemark{a} }&\colhead{$v_{\rm esc,0}$\tablenotemark{b}} & \colhead{$R_h$\tablenotemark{b}}
&\colhead{$L_{\rm bol}$\tablenotemark{a}} &\colhead{$\tau_c$\tablenotemark{c}}\\
\colhead{       }&\colhead{kpc}&\colhead{       }&\colhead{ km~s$^{-1}$     } & \colhead{arcsec} & \colhead{L$_\odot$} & \colhead{yr}
}
\startdata
NGC 104  (47 Tuc)&    4.5   &$-0.76$ &65.9 &190.1 &$1.12\times10^6$&$6\times10^7$\\
NGC 362          &   12.1   &$-1.22$ &46.7 & 49.1 &$4.54\times10^5$&$3\times10^7$\\
NGC 1851         &    8.5   &$-1.16$ &48.7 & 30.4 &$3.88\times10^5$&\nodata\\
NGC 5272 (M3)    &   10.4   &$-1.57$ &27.6 &138.4 &$6.56\times10^5$&$2.1\times10^8$\\
NGC 5904 (M5)	 &    7.5   &$-1.27$ &37.2 & 105.9&$6.21\times10^5$&$2.1\times10^8$\\
NGC 6205 (M13)   &    7.7   &$-1.54$ &32.1 &101.4 &$5.46\times10^5$&$7\times10^7$\\
NGC 6341 (M92)   &    8.2   &$-2.28$ &35.9 & 61.4 &$3.44\times10^5$&$6\times10^7$\\
NGC 6752         &    4.0   &$-1.56$ &\nodata &114.8 &$2.36\times10^5$&\nodata\\
\enddata
\tablenotetext{a}{Distance, metallicity and bolometric luminosity from \citet{harris96}.}
\tablenotetext{b}{Projected half-light radius $R_h$ and central escape velocity $v_{\rm esc,0}$ 
from King-model fits given in \citet{mv05} \citep[except $R_h$ for NGC~6752, from][]{tkd95}.} 
\tablenotetext{c}{Time since last passage through Galactic disk $\tau_c$, from \citet{ode97}.}
\end{deluxetable}

\begin{deluxetable}{lrrrrl}
\tabletypesize{\scriptsize}
\tablecaption{IRAC integrated photometry\label{tab:iracphot}}
\tablewidth{0pt}
\tablehead{
\colhead{Target} &\colhead{$f_{\nu}(3.6)$} & \colhead{ $f_{\nu}(4.5)$} & \colhead{$f_{\nu}(5.8)$} &\colhead{$f_{\nu}(8.0)$}& \colhead{Dataset ID}\\
\colhead{} &\colhead{Jy} &\colhead{Jy} &\colhead{Jy} &\colhead{Jy} &\colhead{}
}
\startdata
NGC 104  &107 & 67.9 & 46.4 & 24.7 & \dataset{ADS/Sa.Spitzer\#0014502656}, \dataset{ADS/Sa.Spitzer\#0007860992}\\
NGC 362  &7.10 & 4.82 & 3.44 & 1.73 & \dataset{ADS/Sa.Spitzer\#0014503168}\\
NGC 1851 &3.57 & 2.31 & 1.64 & 0.82 & \dataset{ADS/Sa.Spitzer\#0014503424}\\
NGC 5272 &8.67 & 5.68 & 3.63 & 1.85 & \dataset{ADS/Sa.Spitzer\#0014504192}\\
NGC 5904 &9.25 & 6.56 & 4.81 & 2.42 & \dataset{ADS/Sa.Spitzer\#0011586304}\\
NGC 6205 &9.66 & 6.43 & 4.35 & 2.17 & \dataset{ADS/Sa.Spitzer\#0014504704}\\
NGC 6341 &5.05 & 3.38 & 2.31 &1.16 & \dataset{ADS/Sa.Spitzer\#14504960}\\
NGC 6752 &13.25 & 8.41 & 6.32 & 3.42 & \dataset{ADS/Sa.Spitzer\#14506240}\\
\enddata
\end{deluxetable}

\begin{deluxetable}{llrrrrl}
\tabletypesize{\scriptsize}
\tablecaption{MIPS cluster photometry\label{tab:mipsphot}}
\tablewidth{0pt}
\tablehead{
\colhead{} &\colhead{}& \multicolumn{2}{c}{half-light} & \multicolumn{2}{c}{core} & \colhead{Dataset IDs}\\
\colhead{Target} &\colhead{Center}&\colhead{$f_{\nu}(24)$} & \colhead{ $f_{\nu}(70)$} & \colhead{$f_{\nu}(24)$} &\colhead{$f_{\nu}(70)$} &\colhead{}\\ 
\colhead{} & \colhead{[J2000.0]} &\colhead{mJy}   &\colhead{mJy} &\colhead{mJy}   &\colhead{mJy} &\colhead{}
}
\startdata
NGC 104  &00:24:07.4 $-$72:04:48&$1980\pm18$ &$124\pm186$&$91\pm2$    &$2\pm9$ & \dataset{ADS/Sa.Spitzer\#0017315328}, \dataset{ADS/Sa.Spitzer\#0017313536}\\ 
NGC 362  &01:03:14.8 $-$70:50:51&$138\pm5$   &$42\pm22$  &$29\pm1.2$  &$5\pm5$ & \dataset{ADS/Sa.Spitzer\#0017316352}, \dataset{ADS/Sa.Spitzer\#0017314560}\\ 
NGC 1851 &05:14:06.8 $-$40:02:49&$63\pm3$   &$-7\pm11$  &$20\pm1$    &$-1\pm4$  &\dataset{ADS/Sa.Spitzer\#0017316096}, \dataset{ADS/Sa.Spitzer\#0017314048} \\ 
NGC 5272 &13:42:11.6 +28:22:47&$161\pm6$  &$1\pm87$   &$10\pm0.7$  &$-2\pm5$  & \dataset{ADS/Sa.Spitzer\#0017316608}, \dataset{ADS/Sa.Spitzer\#0017314816}\\ 
NGC 5904 &15:18:33.0 +02:04:50&$172\pm6$  &$-34\pm67$ &$14\pm0.8$  &$-9\pm5$  & \dataset{ADS/Sa.Spitzer\#0017313024}, \dataset{ADS/Sa.Spitzer\#0017312768}\\
NGC 6205 &16:41:39.9 +36:27:28&$153\pm5$   &$67\pm62$  &$9\pm0.6$   &$-1\pm6$  & \dataset{ADS/Sa.Spitzer\#0017315584}, \dataset{ADS/Sa.Spitzer\#0017313792}\\ 
NGC 6341 &17:17:07.8 +43:08:11&$88\pm4$   &$68\pm25$  &$13\pm0.8$  &$-1\pm5$  & \dataset{ADS/Sa.Spitzer\#0017315840}, \dataset{ADS/Sa.Spitzer\#0017314304}\\ 
NGC 6752 &19:10:52.0 $-$59:59:02&$218\pm6$  &$-5\pm129$ &$13\pm0.8$  &$-4\pm9$ & \dataset{ADS/Sa.Spitzer\#0017315072}, \dataset{ADS/Sa.Spitzer\#0017313280}\\ 
\enddata
\tablecomments{All uncertainties are $1\sigma$.}
\end{deluxetable}

\begin{deluxetable}{lrrrrr}
\tabletypesize{\scriptsize}
\tablecaption{Non-stellar 70\mic fluxes and inferred dust masses\label{tab:ns_dust}}
\tablewidth{0pt}
\tablehead{
\colhead{Target} &\colhead{$f_{\rm 70, tot}$} &\colhead{$f_{\rm 70, core}$} &\colhead{$M_{\rm pred}$} & \colhead{ $M_{\rm total}$} & \colhead{ $M_{\rm core}$}\\
\colhead{} & \colhead{mJy} &\colhead{mJy} & \colhead{M$_\sun$} & \colhead{M$_\sun$} & \colhead{M$_\sun$}
}
\startdata
NGC 104  &$-75\pm186$& $-15\pm9$ &$4.7\times10^{-1}$&$<1.5\times10^{-4}$&$<1.2\times 10^{-5}$\\
NGC 362  &$26\pm22$  & $3\pm5$   &$3.3\times10^{-2}$&$<6.3 \times 10^{-5}$ &$<2.5 \times 10^{-5}$\\
NGC 1851 &$-14\pm11$ & $-6\pm4$  &$>1.1\times10^{-3}$&$<6.4\times 10^{-5}$ &$<4\times 10^{-5}$\\
NGC 5272 &$-15\pm87$& $-6\pm5$   &$1.5\times10^{-1}$&$<3.7\times 10^{-4}$ &$<3.5 \times 10^{-5}$\\
NGC 5904 &$-54\pm67 $& $-18\pm6$&$2.8\times10^{-1}$&$<1.0\times 10^{-4}$ &$<1.5\times 10^{-5}$\\
NGC 6205 &$48\pm62  $& $-4\pm6$ &$4.4\times10^{-2}$&$<1.5\times 10^{-4}$ &$<2.3\times 10^{-5}$ \\
NGC 6341 &$59\pm25  $& $-4\pm5$   &$4.3\times10^{-3}$&$(5.2\pm2.2)\times 10^{-5}$ &$<2.2\times 10^{-5}$\\
NGC 6752 &$-28\pm129$& $-10\pm9$&$>2.6\times10^{-4}$&$<8.2\times 10^{-5}$ &$<9.5\times 10^{-6}$  \\
\enddata
\end{deluxetable}
\end{document}